\documentclass[twocolumn]{jpsj3}
\usepackage{graphicx, color}
\usepackage{bm}
\title{%
Hidden Ladder in SrMoO$_3$/SrTiO$_3$ Superlattices:\\
Experiments and Theoretical Calculations
}

\author{%
Hiroshi Takatsu$^{1,}$\thanks{E-mail address: takatsu@scl.kyoto-u.ac.jp}, 
Naoya Yamashina$^1$, Masayuki Ochi$^2$, 
Hsin-Hui Huang$^3$, Shunsuke Kobayashi$^3$, 
Akihide Kuwabara$^3$, 
Takahito Terashima$^4$, 
Kazuhiko Kuroki$^2$, and 
Hiroshi Kageyama$^1$\thanks{E-mail address: kage@scl.kyoto-u.ac.jp}
}

\inst{%
$^1$Department of Energy and Hydrocarbon Chemistry, Graduate School of Engineering, Kyoto University, Kyoto 615-8510, Japan\\
$^2$Department of Physics, Osaka University, Machikaneyama-cho, Toyonaka, Osaka 560-0043, Japan\\
$^3$Nanostructures Research Laboratory, Japan Fine Ceramics Center, Atsuta, Nagoya 456-8587, Japan\\
$^4$Department of Physics, Graduate School of Science, Kyoto University, Kyoto 606-8502, Japan
}

\recdate{\today}

\abst{%
A double-layered perovskite oxide Sr$_3$Mo$_2$O$_7$ is considered 
a ``hidden ladder'' system with wide and narrow bands near the Fermi level, 
for which high-$T_{\rm c}$ superconductivity is expected. 
However, the difficulty in synthesis, especially in the preparation of samples
without oxygen deficiency, can hinder the observation of superconductivity. 
In this study, we constructed a double-layer SrMoO$_3$ block through artificial
superlattices with the insulating SrTiO$_3$ block, 
(SrMoO$_3$)$_m$/(SrTiO$_3$)$_t$ ($m = 2, 4$; $t = 4$). 
First-principles calculations for bilayered SrMoO$_3$ ($m = 2$) exhibit 
a wide-narrow band structure near the Fermi level, which bears 
a close resemblance to Sr$_3$Mo$_2$O$_7$. 
The dispersion along the $k_z$ direction is strongly suppressed by increasing 
the number of the SrTiO$_3$ layers, $t$. 
However, no superconductivity is observed down to 0.1~K. 
We discuss the absence of the superconductivity for the present films on 
the basis of results of scanning transmission electron microscopy and 
band structure calculations.
}

\begin{document}
\maketitle

\section{Introduction}
There has been a rapid progress in the study of unconventional 
iron-based superconductors~\cite{HosonoPhysica2015,Chubukov2015}.
Of particular interest is the role of an incipient band - a band that 
does not intersect but lies close to the Fermi level. 
The incipient band is expected to enhance superconductivity, 
as suggested in heavily electron doped $A_x$Fe$_{2-y}$Se$_2$ 
($A=$ alkali metal ions)~\cite{GuoPRB2010,FangEPL2011} and 
monolayer FeSe~\cite{WangCPL2012,LiuNC2012,GeNM2015,ShiogaiNP2016}.
Superconducting mechanisms have been discussed using finite-energy 
spin fluctuations~\cite{Hirschfeld2011,Wang2011,Bang2014,ChenPRB2015,Bang2016,MishraSR2016,MishraSR2019}.
The incipient-band induced high-$T_{\rm c}$ superconductivity was 
initially addressed 
in 2005 for cuprates with a ladder-type structure~\cite{KurokiPRB2005}.
For the two-leg ladder model with bonding and antibonding bands, one of 
the bands becomes wide and the other becomes narrow (or flat) once 
the diagonal hopping is introduced, and an interband pair-scattering 
process results in high-$T_{\rm c}$ superconductivity.
A similar idea has been extended to various quasi-one-dimensional 
lattice models~\cite{MatsumotoPRB2018} as well as 
other lattice models~\cite{KobayashiPRB2016,NakataPRB2017,MisumiPRB2017}.
%

\begin{figure}[t]
\begin{center}
 \includegraphics[width=0.45\textwidth]{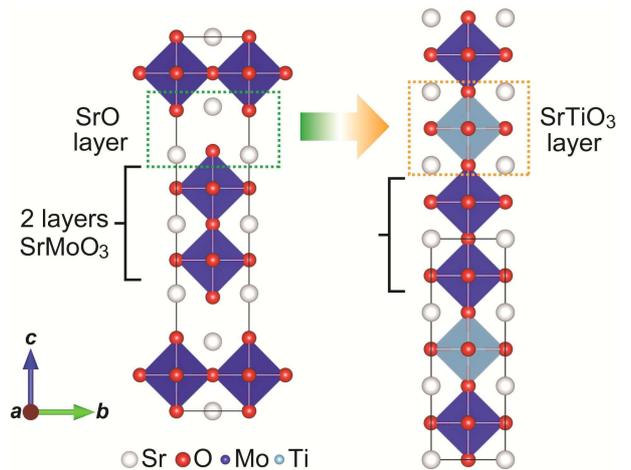}
\caption{
(Color online) 
Crystal structures of Sr$_3$Mo$_2$O$_7$ and the artificial superlattice of
(SrMoO$_3$)$_m$/(SrTiO$_3$)$_t$ for $(m,t) = (2,1)$.
}
\label{fig.1}
\end{center}
\end{figure}
%
Recently, Ogura {\it et al} have proposed a new type of materials that possess
wide and incipient-narrow bands, bilayer Ruddlesden-Popper (RP) compounds, 
Sr$_3$Mo$_2$O$_7$ and Sr$_3$Cr$_2$O$_7$ (Mo$^{4+}$, Cr$^{4+}$, $d^2$) ~\cite{OguraPRB2017}.
Although the bilayer RP structures are apparently not ladder-shaped  (Fig.~\ref{fig.1}), 
they have electronic structures similar to the two-leg ladder cuprate.
This results from the anisotropy of the $d_{xz}$ and $d_{yz}$ orbitals;
the $d_{xz}$ and $d_{yz}$ orbitals can broadly be treated separately and 
form the two-leg ladders along the $x$ and $y$ directions,
respectively~\cite{OguraPRB2017}.
Theoretical calculations on the basis of the fluctuation exchange 
approximation have shown the potential for high-$T_{\rm c}$ 
superconductivity in this ``hidden ladder'' system~\cite{OguraPRB2017}.

Unfortunately, no sign of superconductivity has been observed experimentally in 
Sr$_3$Cr$_2$O$_7$ and Sr$_3$Mo$_2$O$_7$~\cite{JeanneauPRL2017,KounoJPSJ2007},
although the incipient band is located close to the Fermi level.
For Sr$_3$Cr$_2$O$_7$, orbital ordering accompanied by an antiferromagnetic order
at 210~K may prevent superconductivity~\cite{JeanneauPRL2017}, while in 
the case of Sr$_3$Mo$_2$O$_7$, oxygen deficiency is likely to be the main 
cause~\cite{OguraPRB2017}.
For example, Sr$_3$Mo$_2$O$_7$ grown under high pressure has 
the oxygen content of $\sim6.3$~\cite{OguraPRB2017}. 
Some efforts have been made to control of the partial pressure of 
oxygen~\cite{Steiner1998,KounoJPSJ2007}, but it appears there remains 
ambiguity in the chemical composition.

Epitaxial thin film growth has an advantage for new materials synthesis in 
terms of rational design of materials toward achieving desired properties~\cite{KosterBOOK2015,ArthurSS2002,Gorbenko2002,SchlomJACS2008,OkaCrystEngComm2017}.
More specifically, artificial heterostructures of functional oxides can be used
as a platform to tailor their functionalities as one can modify multiple degrees
of coupling between lattices, electrons, and spins and offer a controlled dimensionality. 
In this paper, we fabricated artificial superlattices of SrMoO$_3$ and SrTiO$_3$,
(SrMoO$_3$)$_m$/(SrTiO$_3$)$_t$ ($m = 2, 4$; $t = 4$), with a primary motivation
to realize the hidden ladder, which is the case for a double-SrMoO$_3$ block 
($m = 2$) (Fig.~\ref{fig.1}), where the SrTiO$_3$ block serves as an insulating
block layer, instead of the SrO layers in Sr$_3$Mo$_2$O$_7$.
We show physical properties of the obtained films with $m = 2$ and 4, 
as well as theoretical calculation of these artificial superlattices. 

\section{Calculations and Experiments}
First-principles band structure calculation for 
(SrMoO$_3$)$_m$/(SrTiO$_3$)$_t$ superlattices
were performed using the WIEN2k package~\cite{WIEN2k}. 
We used the Perdew-Burke-Ernzerhof parameterization of 
the generalized gradient approximation~\cite{PerdewPRL1996}, 
with $RK_{\rm max} = 7$ and $12\times12\times k_{z}'$ 
($k_{z}' = 12/t+m$ for $t+m\le4$ and $k_{z}' =2$ for $4<t+m$). 
For these calculations, we considered a periodic structure model 
(i.e., a layered oxide (SrMoO$_3$)$_m$(SrTiO$_3$)$_t$
with the space group of $P4/mmm$ (No.123) as shown in Fig.~\ref{fig.1}),
assuming the ideal case of no long range orders,
and optimized the structural parameters using the QUANTUM ESPRESSO
package~\cite{GiannozziJPCM2009,GiannozziJPCM2017}.

Epitaxial films of the (SrMoO$_3$)$_m$/(SrTiO$_3$)$_t$ superlattice
were grown on the (001)-oriented SrTiO$_3$ or KTaO$_3$ substrate using
a custom-made reactive molecular beam epitaxy (MBE) system 
(EGL-1420-E2, Biemtron). 
The films were deposited at a substrate temperature of 520~$^\circ$C under 
O$_2$ gas flow with a background pressure of about $4\times10^{-7}$~Torr.
The present optimal condition was employed in the previous study on 
SrMoO$_3$ films with MBE~\cite{Takatsu_growth_SrMoO$_3$}. 
The surface structure of the film and substrate was monitored 
{\it{in}-\it{situ}} by reflection high-energy electron diffraction (RHEED) 
with an acceleration voltage of 20~keV.
Superlattices were synthesized in the following way: 
firstly, we estimated average interval of oscillation peaks of RHEED 
intensity of SrMoO$_3$ and SrTiO$_3$ 
at the beginning of the growth, and then we used these durations for 
the growth of the superlattice film~\cite{supplement_SMO_SL}. 
The RHEED intensity oscillation was observed during 
the growth of both films,
suggesting that the films were epitaxially grown in 
a lateral mode~\cite{KosterBOOK2015,ArthurSS2002},
although the intensity decreases as the synthesis of the superlattice 
proceeds.
%
\begin{figure*}[t]
\begin{center}
 \includegraphics[width=0.92\textwidth]{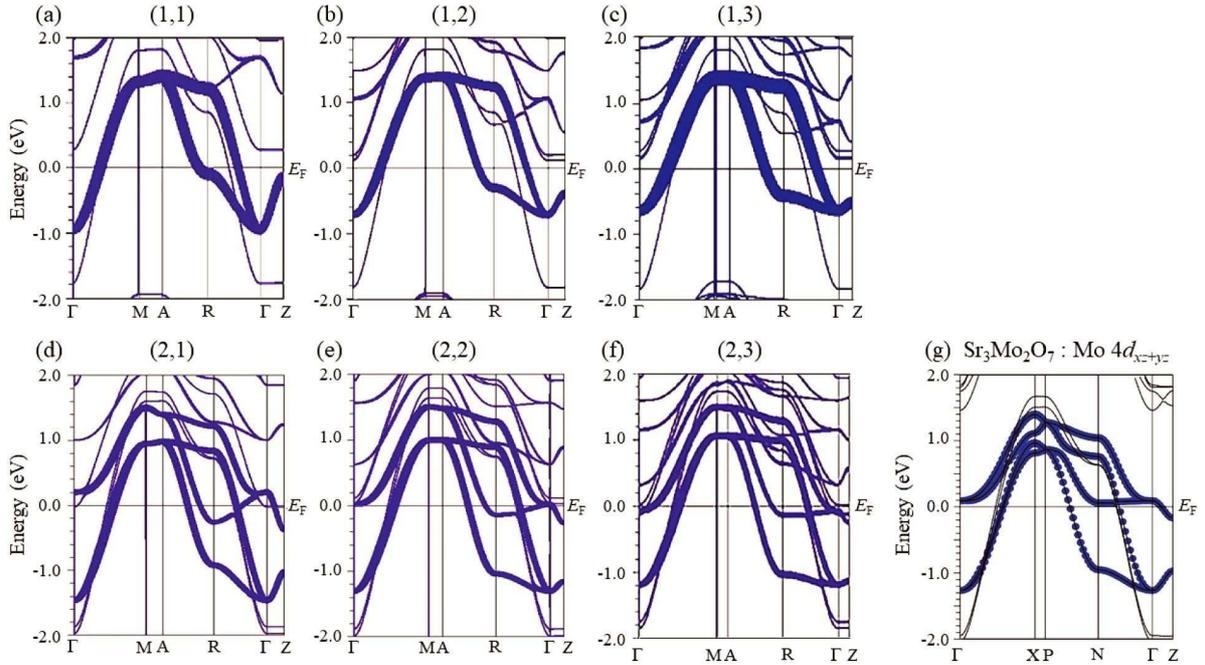}
\caption{
(Color online) 
Band dispersions of (a)--(f) (SrMoO$_3$)$_m$/(SrTiO$_3$)$_t$ 
for $(m,t)=(1,1)$--$(1,3)$, and $(m,t)=(2,1)$--$(2,3)$
and (g) Sr$_3$Mo$_2$O$_7$, 
where contributions of the Mo-$d_{xz}$ and $d_{yz}$ orbitals are highlighted
to emphasize the ladder-like band.
Note that symmetric points of M, A and R of (SrMoO$_3$)$_m$/(SrTiO$_3$)$_t$ 
with the $P4/mmm$ space group correspond to X, P, and N of Sr$_3$Mo$_2$O$_7$ with the $I4/mmm$ space group, respectively.
}
\label{fig.2}
\end{center}
\end{figure*}

%
X-ray diffraction (XRD) measurements after the growth 
were carried out at room temperature (RT) with  
a Rigaku SmartLab diffractometer equipped with a Cu K$\alpha_1$ monochromator.
The XRD data were analyzed by using the GlobalFit software installed in 
the Rigaku SmartLab XRD system. 
Complementary to XRD, the obtained superlattice film was also characterized 
by an aberration-corrected 
scanning transmission electron microscopy (STEM, JEM-2400FCS, JEOL Ltd.). 
STEM observations were performed at accelerating voltages of 200 kV. 
The probe-forming aperture semiangle was 22~mrad and high-angle annular
dark-field (HAADF) STEM images were recorded with 68--280~mrad detectors.
The cross-sectional STEM sample was prepared by dual-beam focused ion beam 
scanning microscopy (NB5000, Hitachi High-Technologies Co.) using Ga ions. 
After this treatment, 
Ar ion milling with a cold stage was performed.
The electrical resistivity $\rho$ was measured with a standard 
four-probe method. The samples were cooled down to 0.1~K using 
an adiabatic demagnetization refrigerator installed in 
a Quantum Design physical property measurement system (PPMS). 
\section{Results and discussion}
\subsection{Band structures for the (SrMoO$_3$)$_m$/(SrTiO$_3$)$_t$ superlattice}
In Fig.~\ref{fig.2}, we show first-principles electronic band structures of 
the (SrMoO$_3$)$_m$/(SrTiO$_3$)$_t$ superlattice with a single-SrMoO$_3$ layer
($m = 1$) or a double- SrMoO$_3$ layers ($m = 2$) and insulating SrTiO$_3$ 
layers of different thickness ($1\le t \le 3$), where contributions of 
the Mo-$d_{xz}$ and $d_{yz}$ orbitals are highlighted by thick lines. 
In the $m=1$ system, the Mo-$d_{xz}$ and $d_{yz}$ orbitals form the chain-like
networks along the $x$ and $y$ directions, respectively, which results from 
orbital anisotropy and the presence of insulating SrTiO$_3$ layers.
The Mo-$d_{xz/yz}$ band structures for the $m=1$ system shown in
Figs.~\ref{fig.2}(a)--(c) look similar. 
An important difference, however, is the $\Gamma$--Z dispersion 
(i.e., the dispersion along the $k_z$ direction), which corresponds to 
the coupling among SrMoO$_3$ layers separated by SrTiO$_3$ layers. 
By increasing the number of the SrTiO$_3$ layers ($t$), 
such coupling is rapidly weakened.

In the $m=2$ system, the Mo-$d_{xz}$ and $d_{yz}$ orbitals in the SrMoO$_3$
bilayer form the two-leg ladders along the $x$ and $y$ directions, respectively.
As a result, the Mo-$d_{xz/yz}$ band dispersion consists basically of 
the bonding and anti-bonding bands, each of which has a shape similar to 
the Mo-$d_{xz/yz}$ bands in the $m=1$ system. Here, we use the terms of bonding
and anti-bonding to denote the coupling between the SrMoO$_3$ layers along 
the $z$ direction. Since these two sets of the bonding and anti-bonding bands
have a different bandwidth as shown in Figs.~\ref{fig.2}(d)--(f), 
one can say that the wide and narrow bands coexist. 
The same situation occurs in Sr$_3$Mo$_2$O$_7$, where the $d_{xz/yz}$ hidden
ladders and the resulting coexistence of the wide and narrow bands were
theoretically pointed out~\cite{OguraPRB2017}.
One can also see that the narrow band edge is close to the Fermi energy 
$E_{\rm F}$, 
which results in the incipience of narrow bands similarly to Sr$_3$Mo$_2$O$_7$.
Because the sandwiched SrTiO$_3$ layers are insulating, 
the Mo-$d_{xz/yz}$ band structures in the $m=2$ system resemble 
that for Sr$_3$Mo$_2$O$_7$ shown in Fig.~\ref{fig.2}(g). 
As in the $m=1$ system, the $\Gamma$--Z dispersion in 
the $m=2$ system becomes weak with increasing $t$. 
In fact, a weaker dispersion is found in the $(m,t)=(2,3)$ case 
than Sr$_3$Mo$_2$O$_7$, suggesting that the (SrMoO$_3$)$_m$/(SrTiO$_3$)$_t$ 
superlattice with a sufficiently large $t$ is favorable for superconductivity.
Thus, (SrMoO$_3$)$_m$/(SrTiO$_3$)$_t$ superlattices with $m=2$ are regarded 
as hidden two-leg ladder system that can host high-$T_{\rm c}$ superconductivity
even without carrier doping, as suggested by the many-body analysis of 
Sr$_3$Mo$_2$O$_7$~\cite{OguraPRB2017}.
The observations shown here are somewhat reminiscent of 
the quantum well states in superlattices or ultra-thin
films~\cite{YoshimatsuScience2011,BorisScience2011,EsakiIBM1970}. 

\subsection{Experiments on $(m,t)=(2,4)$ superlattice films}
We initially synthesized the superlattice films of double SrMoO$_3$ layers 
($m = 2$) combined with quadruple SrTiO$_3$ layers ($t = 4$)
on a (001)-oriented SrTiO$_3$ substrate.
Figure~\ref{fig.3}(a) shows the out-of-plane $\theta$--$2\theta$ XRD pattern
for the film containing eleven repetitions of the $(m,t)=(2,4)$ structure,
11[(SrMoO$_3$)$_2$/(SrTiO$_3$)$_4$]. 
This film exhibits distinct satellite peaks, indicating the formation of 
the superlattice, 
{\color[rgb]{0,0,0} although a tiny unknown peak (probably due to 
MoO$_3$ related secondary phase~\cite{Magneli1948,M.SatoJPC1987,PariseJSSC1991}) 
is present at about $25.5^\circ$. The experimental diffraction pattern is also 
deviated slightly from the calculation,} 
based on dynamical theory of X-ray
diffraction~\cite{TakahashiSS1995,YashiroSS2001},
for the ideal structure of 11[(SrMoO$_3$)$_2$/(SrTiO$_3$)$_4$]
(the blue solid line in Fig.\ref{fig.3}(a)).

Figure~\ref{fig.3}(b) shows the temperature dependence of the electrical
resistivity $\rho$ of the same sample.
No superconductivity was observed at temperatures down to 2~K.
While $\rho$ is almost temperature independent, below 100~K it 
increases with decreasing temperature.
Since SrTiO$_3$ and SrMoO$_3$ are both magnetically inactive,
this increase of $\rho$ at low temperatures is not due to 
the Kondo effect~\cite{KondoPTP1964}, but is ascribable to 
disordered-caused transport anomalies
such as variable range hopping (VRH)~\cite{MottPM1968}, 
weak Anderson localization (WAL)~\cite{AbrahamsPRL1979,AndersonPRL1979}, 
and impurity-induced electron-electron (IIEE) interaction~\cite{AltshulerPRL1980},
which have been observed also in films and wires~\cite{LeeRMP1985}. 
{\color[rgb]{0,0,0} 
Owning to insulating $d_0$ nature and large $\rho$ of the Mo(VI)O$_3$ related impurity phases, 
one can ignore their contribution to the temperature dependence of $\rho$.} 
Instead, 
the present result may be related to a certain disorder in the interface between 
SrTiO$_3$ and SrMoO$_3$, which could cause the discrepancy in 
XRD intensity between experiment and calculation, and also account for 
the absence of superconductivity since superconductivity mediated by 
spin fluctuations 
tends to be affected by disorders or impurities involved in the main phase~\cite{TarasconPRB1987}.
The interface roughening or Ti/Mo intermixing disorder was found in STEM for 
the (4,4) film, which will be discussed later.
{\color[rgb]{0,0,0} Note that thick films of SrMoO$_3$, synthesized by 
the same MBE method, showed an improved $\rho$ of 24~$\mu\Omega$~cm at RT, 
as compared with the thick films by other techniques~\cite{Inukai1985,WangJCG2001,Radetinac2010,Radetinac2016}.
On the other hand, a slight influence of extrinsic effects such as Mo(VI)-containing impurities is still 
seen in $\rho$~\cite{Takatsu_growth_SrMoO$_3$}.
We suspect the emergence of such impurities affects the crystallinity of 
the heterointerface of the present superlattices, as discussed in a later section.
}

\subsection{Experiments on $(m,t)=(4,4)$ superlattice films}
\begin{figure}[t]
\begin{center}
 \includegraphics[width=0.45\textwidth]{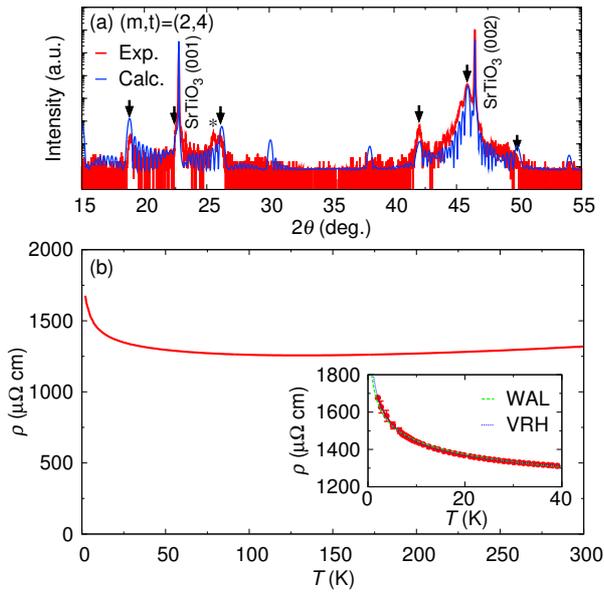}
\caption{
(Color online) 
(a) Out-of-plane $\theta$--2$\theta$ XRD patterns (red line) of 
a (SrMoO$_3$)$_m$/(SrTiO$_3$)$_t$ superlattice film on 
the (001)-oriented SrTiO$_3$ substrate for $(m,t) = (2,4)$.
The blue line represents a dynamical theory diffraction calculation.
{\color[rgb]{0,0,0} The fundamental and satellite peaks observed 
are indicated by arrows.
An asterisk $\ast$ represents unknown peak.}
(b) Temperature dependence of the electrical 
resistivity $\rho$ of the same sample as (a).
The inset shows the magnified view of the data plot below $T=40$~K,
with fitting curves of WAL ($\rho\propto \log(T)$) and VLH in 2D 
($\rho\propto \exp(T_0/T)^{1/d+1}$ with $d=2$).
Note that the IIEE interaction also brings 
about logarithmic temperature dependence~\cite{LeeRMP1985, AltshulerPRL1980}. 
These fittings imply a large influence of chemical disorder on resistivity.
}
\label{fig.3}
\end{center}
\end{figure}
\begin{figure}[t]
\begin{center}
 \includegraphics[width=0.40\textwidth]{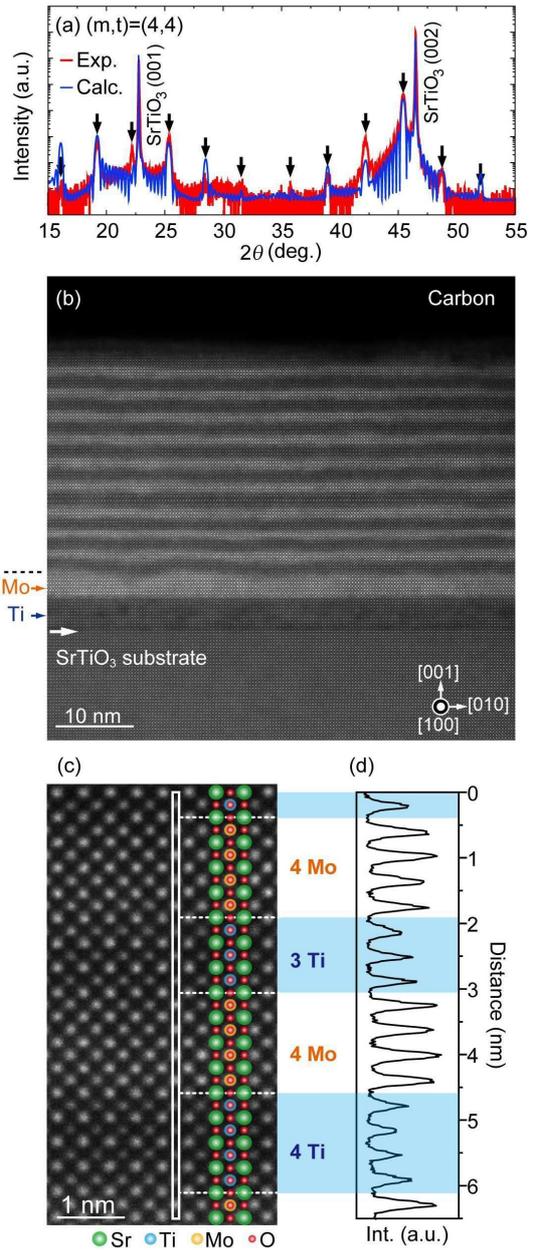}
\caption{
(Color online) 
(a) Out-of-plane $\theta$--2$\theta$ XRD patterns of 
a (SrMoO$_3$)$_m$/(SrTiO$_3$)$_t$ superlattice film on 
the (001)-oriented SrTiO$_3$ substrate for $(m,t) = (4,4)$.
The blue line represents a dynamical theory diffraction calculation.
{\color[rgb]{0,0,0} The fundamental and satellite peaks are indicated by arrows.
A peak at about $22.5^\circ$ is a superlattice peak, not due to an 
impurity secondary phase, although it is slightly deviated from the calculation.}
(b) HAADF STEM image of the SrMoO$_3$/SrTiO$_3$ superlattice film taken from 
the [100] zone axis of the SrTiO$3$ substrate. 
The same sample as (a) was used for this experiment.
The white arrow indicates the interface between substrate and superlattice film. 
(c) Magnified view of the HAADF STEM image of the superlattice and 
(d) the intensity profile of $B$-site cation (Ti or Mo) columns 
obtained from the white rectangle region in (c).
}
\label{fig.4}
\end{center}
\end{figure}
\begin{figure}[t]
\begin{center}
 \includegraphics[width=0.45\textwidth]{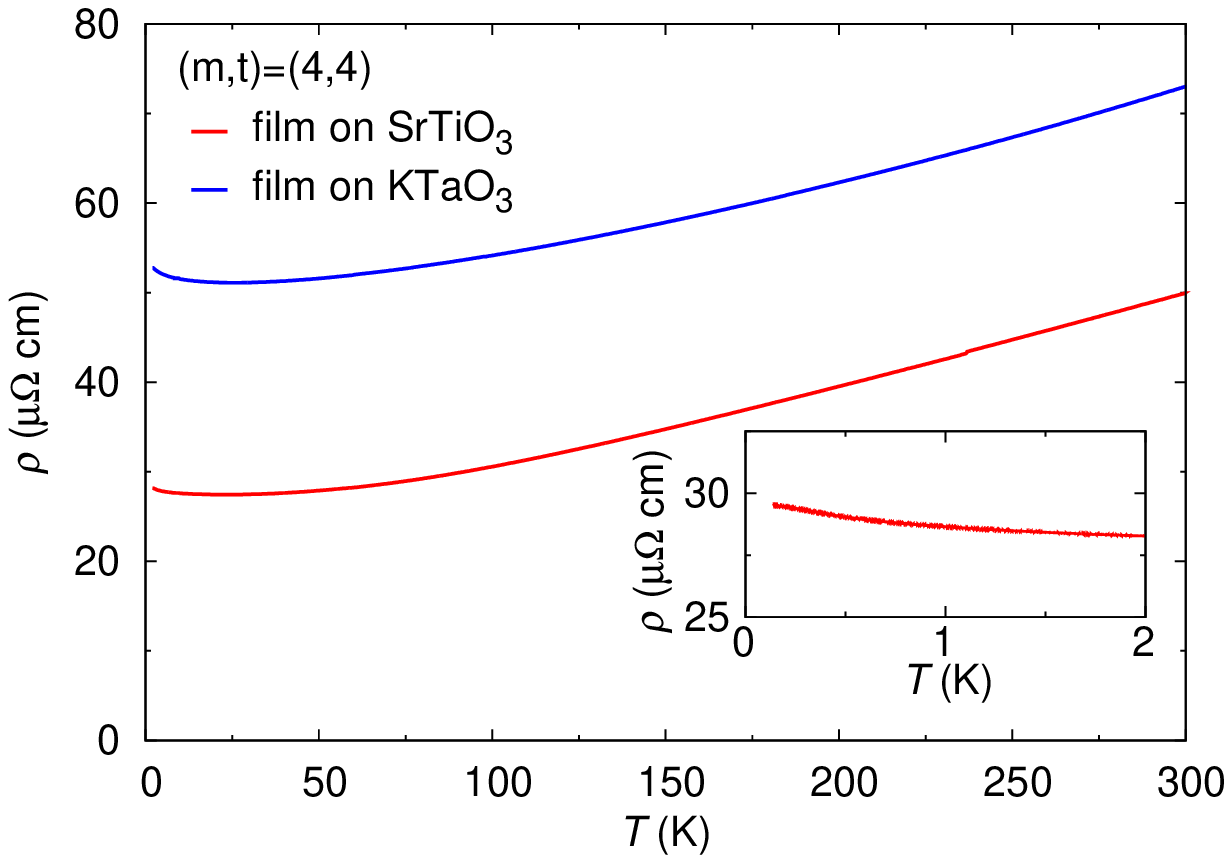}
\caption{
(Color online) 
Temperature dependence of the electrical 
resistivity $\rho$ of the same film as Fig.~\ref{fig.4}(a) 
on the (001)-oriented SrTiO$_3$ substrate 
and the $(4,4)$ film on the (001)-oriented KTaO$_3$ substrate.
The slight enhancement of $\rho$ in the film on the KTaO$_3$ substrate 
could be attributed to enhanced impurity scattering around roughened interfaces.
The inset shows the low-temperature behavior of $\rho$ of 
the film on the SrTiO$_3$ substrate.
}
\label{fig.5}
\end{center}
\end{figure}
%
We examined several growth conditions for the double-layer SrMoO$_3$ ($m = 2$)
films, but failed to obtain a superlattice film free from a chemical disorder
around the heterointerface. Accordingly, we synthesized a superlattice film with
$(m,t) = (4,4)$, with a hope that the inner two SrMoO$_3$ layers might not have
chemical disorder.
{\color[rgb]{0,0,0} Although 
the heterointerface of the film might be disordered inevitably,
we expected that such an inner double-SrMoO$_3$-layer
structure works as the hidden ladder lattice and may lead to superconductivity.}

Figure~\ref{fig.4}(a) shows the out-of-plane $\theta$--$2\theta$ XRD 
pattern for the film having ten repetitions of the $(m,t)=(4,4)$ structure,
10[(SrMoO$_3$)$_4$/(SrTiO$_3$)$_4$], grown on 
the (001)-oriented SrTiO$_3$ substrate. 
Fundamental and superlattice satellite peaks are clearly seen.
The simulated pattern for the $(4,4)$ structure broadly agrees with 
the experimental data. 
{\color[rgb]{0,0,0} The same superlattice structure was also
constructed on the (001)-oriented KTaO$_3$ substrate with 
a similar film quality~\cite{supplement_SMO_SL}.} 
It is worth noting that the $(4,4)$ film has better
quality than the $(2,4)$ film, which can be seen from a better match  
between the experimental and theoretical patterns for the former film 
(Figs.~\ref{fig.3}(a) and \ref{fig.4}(a)).

STEM observations for the $(4,4)$ film provide detailed information on the structure, 
with interface roughening (Fig.~\ref{fig.4}(b)) and intermixing disorder
(Figs.~\ref{fig.4}(c)-(d)) around the interface. 
{\color[rgb]{0,0,0} We used the film fabricated on the SrTiO$_3$ substrate for these experiments, 
since $\rho$ is slightly lower (i.e., the quality of the film is slightly higher) than 
that of the film on the KTaO$_3$ substrate, as discussed in a later section.}
The roughness and site exchange around the interface between SrMoO$_3$ and
SrTiO$_3$ films would primarily be ascribed to a poorer crystallinity of
SrMoO$_3$ vs. SrTiO$_3$.
An insulating Mo(VI)-containing impurity, which often appears in SrMoO$_3$ 
films~\cite{WadatiPRB2014,Radetinac2016,Salg2019},
may affect the crystallinity of SrMoO$_3$.
STEM measurements revealed trace impurity with a structure different 
from perovskites~\cite{supplement_SMO_SL}.
Interface roughening could also result from the lattice mismatch between
SrMoO$_3$ ($a=3.976$~\AA)~\cite{MacquartJSSC2010} and 
SrTiO$_3$ ($a=3.905$~\AA)~\cite{Schmidbauer2012}.
Additionally, the difference in surface energies between SrMoO$_3$ and SrTiO$_3$ 
films may be at play, which has been suggested in EuMoO$_3$/SrTiO$_3$
superlattices~\cite{FujitaPRB2013}.

Although the heterointerface of the $(4,4)$ film is disordered, the inner two
layers of the quadruple SrMoO$_3$ layers look clean (Figs.~\ref{fig.4}(b)-(c)).
Using this film, we measured the temperature dependence of $\rho$, 
the result of which is shown in Fig.~\ref{fig.5}, along with 
a 10[(SrMoO$_3$)$_4$/(SrTiO$_3$)$_4$] film on the KTaO$_3$ substrate.
Both films exhibit metallic temperature dependence with a positive slope
($d\rho/dT>0$), as opposed to the $(2,4)$ film (Fig.~\ref{fig.3}(b)).
{\color[rgb]{0,0,0} The slight enhancement of $\rho$ is seen in the film on 
the KTaO$_3$ substrate, probably due to the enhancement of impurity scattering. 
There may be some effect of the epitaxial tensile strain from the KTaO$_3$ substrate, 
the actual effect of which is not clear in physical properties of the film obtained.
On the other hand, 
it is noteworthy that $\rho$ of both films on the SrTiO$_3$ and 
KTaO$_3$ substrates is much lower (about 30 times smaller) than 
that of the $(2,4)$ film.} 
This result implies that charge carriers in 
the $(4,4)$ film mainly move through the inner bilayers;
{\color[rgb]{0,0,0} i.e., the influence of impurities or disorder 
is small in the inner bilayers.
However, no superconductivity nor any signature of magnetic phase 
transitions were observed down to 0.1~K 
(the inset of Fig.~\ref{fig.5}).}

\subsection{Band structure calculation for the $(m,t)=(4,4)$ superlattice and the absence of superconductivity}
\begin{figure}[t]
\begin{center}
 \includegraphics[width=0.40\textwidth]{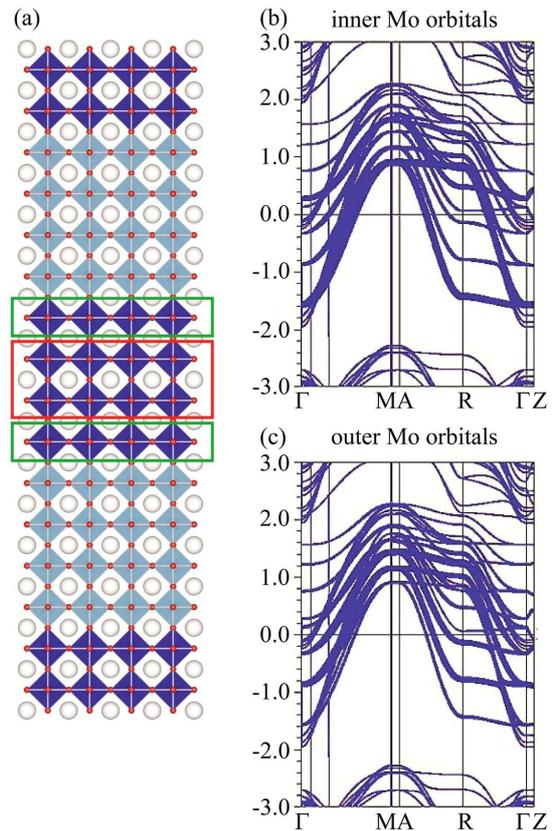}
\caption{
(Color online) 
(a) Crystal structure of (SrMoO$_3$)$_4$/(SrTiO$_3$)$_4$ and 
{\color[rgb]{0,0,0} 
(b,c) its band structure, where the Mo-$d_{xz}$/$d_{yz}$ orbital 
weight in the inner and outer layers is highlighted, respectively.
The inner and outer SrMoO$_3$ layers are shown by 
the red and green squares in (a), respectively.}
}
\label{fig.6}
\end{center}
\end{figure}
%
We performed first-principles band structure calculations for the disorder-free
$(4,4)$ superlattice. Here, we separately examine the contributions from 
the inner and outer bilayers (Fig.~\ref{fig.6}(a)).
Figures 6(b) and (c) show the band structure, 
{\color[rgb]{0,0,0} where contributions of 
the Mo-$d_{xz}$/$d_{yz}$ orbitals in the inner and outer layers are
highlighted by thick lines, respectively.}
Notably, the narrow band of the inner bilayers
is far away from $E_{\rm F}$ (Fig.~\ref{fig.6}(b)) and hence may not function 
as an incipient band that contributes to the enhanced superconductivity.
On the other hand, the band of the outer layers sits close to $E_{\rm F}$
(Fig.~\ref{fig.6}(c)) and can work as an incipient band. 
It is thus possible that the superconducting correlation is suppressed in 
the superlattice with $m = 4$ in the presence of disorder at the heterointerface. 
Oxygen deficiency, discussed in the previous theoretical study~\cite{OguraPRB2017},
is also a possible origin for the absence of superconductivity.

Although superconductivity was not observed, this study has clearly demonstrated
that electronic structures similar to a naturally available Sr$_3$Mo$_2$O$_7$ 
(and cuprate ladders) can be obtained simply by constructing artificial
superlattice of perovskite SrMoO$_3$ and SrTiO$_3$, with a further room 
for modification of electronic structures by changing a repetition of 
stacking sequence. 
The present system is in a sense analogous to the quantum well structure 
of perovskites studied in ultra-thin films~\cite{YoshimatsuScience2011}; 
SrVO$_3$ with the $d^1$ configuration exhibits a metal insulator transition at 
a critical thickness of 2--3 layers through a pseudogap region with a thickness
below 6 layers~\cite{YoshimatsuPRL2010}.
Thus electron doping to the insulator phase could be achieved by using 
SrMoO$_3$ with the $d^2$ configuration.

\section{Conclusion}
We have fabricated the superlattice of SrMoO$_3$ and SrTiO$_3$,
(SrMoO$_3$)$_m$/(SrTiO$_3$)$_t$, inspired by the recent theoretical 
prediction of superconductivity in the bilayer RP compound Sr$_3$Mo$_2$O$_7$. 
First-principles calculations for (SrMoO$_3$)$_2$(SrTiO$_3$)$_t$ ($t\ge2$) 
show similar characteristics with Sr$_3$Mo$_2$O$_7$, which features 
a ladder-like electronic band structure with wide and narrow bands, 
suggesting a potential high-$T_{\rm c}$ superconductivity in this artificial
superlattice. Experimentally, we have succeeded in growing the superlattices
with $(m,t) = (2,4)$ and $(4,4)$.
The absence of superconductivity in our films might be due to a chemical
disorder in the interface. Further optimization of growth conditions is
essential to reduce interface roughening and/or intermixing disorder. 
Such a superlattice with SrMoO$_3$ is also an interesting future topic 
as quantum well structures.

\section*{Acknowledgment}
This work was supported by CREST (JPMJCR1421) and JSPS KAKENHI 
Grants (No.~JP16H06439, No.~JP16H06440, No.~JP17H04849, No.~JP17H05481, 
No.~JP18H01860, No.~JP18K13470, and No.~JP19H04697).

\bibliography{69641_with_SI.bbl}

\clearpage
\onecolumn
\begin{center}
\textbf{\large Supplements for ``Hidden Ladder in SrMoO$_3$/SrTiO$_3$ Superlattices: Experiments and Theoretical Calculations''}
\end{center}
\setcounter{equation}{0}
\setcounter{figure}{0}
\setcounter{table}{0}
\setcounter{page}{1}
\renewcommand{\theequation}{S\arabic{equation}}
\renewcommand{\thefigure}{S\arabic{figure}}


\section*{Abstract}
{\color[rgb]{0,0,0}In this supplemental material, we show experimental results of 
the high-energy electron diffraction (RHEED) intensity oscillation 
during the growth of a superlattice film,
X-ray diffraction (XRD) data of a (SrMoO$_3$)$_4$/(SrTiO$_3$)$_4$ film on 
the KTaO$_3$ substrate, additional data of 
scanning transmission electron microscopy (STEM) for a superlattice film.
}

\section*{RHEED intensity oscillation}
\begin{figure}[h]
\begin{center}
 \includegraphics[width=0.55\textwidth]{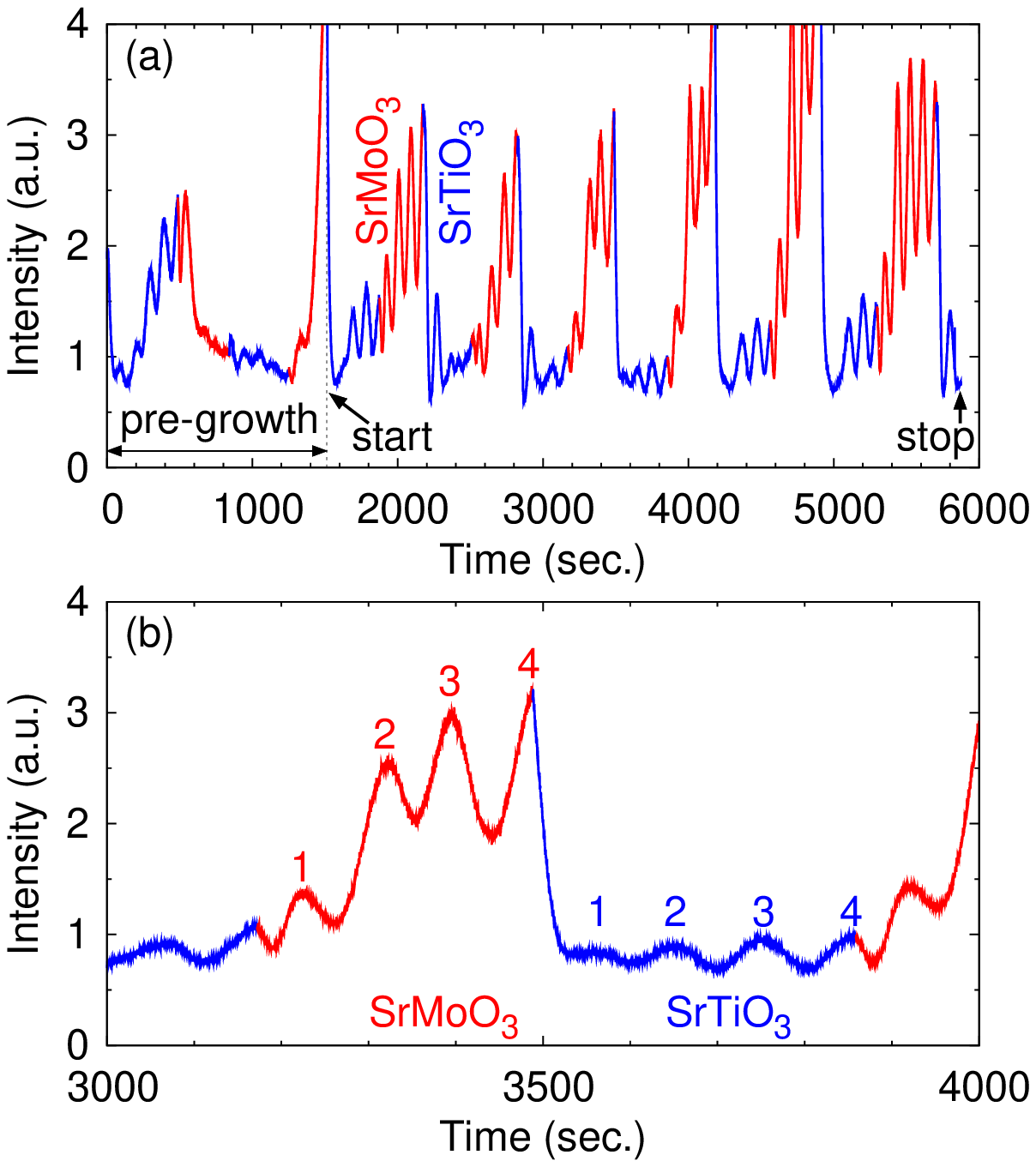}
\caption{
(a) RHEED intensity oscillations during the growth of (SrMoO$_3$)$_m$/(SrTiO$_3$)$_t$ with $(m,t)=(4,4)$.
(b) Magnified view of the intensity of the RHEED oscillation, shown in (a).
}
\label{fig.s1}
\end{center}
\end{figure}

Figure~\ref{fig.s1} shows one of the typical results of 
RHEED intensity oscillation during the growth of
superlattice film of (SrMoO$_3$)$_4$/(SrTiO$_3$)$_4$ on the (001)-oriented
SrTiO$_3$ substrate. 
%
At the beginning of the film growth (pre-growth), 
we estimated the average time of 
the interval of oscillation peaks, since
the interval corresponds to the growth of one unit cell of SrMoO$_3$ 
(or  SrTiO$_3$)~\cite{TerashimaPRL1990}.
We often observed that the RHEED intensity decreases when switching from 
the growth of SrTiO$_3$ to that of SrMoO$_3$. 
This behavior is probably due to the emergence of interface roughening or 
disorder around the heterointerface, both of which were detected by STEM experiments.
However, the intensity oscillations can be maintained to the end of the growth.
Therefore, we utilized this behavior for the growth of the superlattice, i.e.,
each layer thickness was controlled by {\it in-site} monitoring of 
the intensity oscillation in addition to the use of the averaged time 
of the oscillation interval.
%

\section*{XRD results of a (SrMoO$_3$)$_4$/(SrTiO$_3$)$_4$ film on the KTaO$_3$ substrate}
\begin{figure}[h]
\begin{center}
 \includegraphics[width=0.55\textwidth]{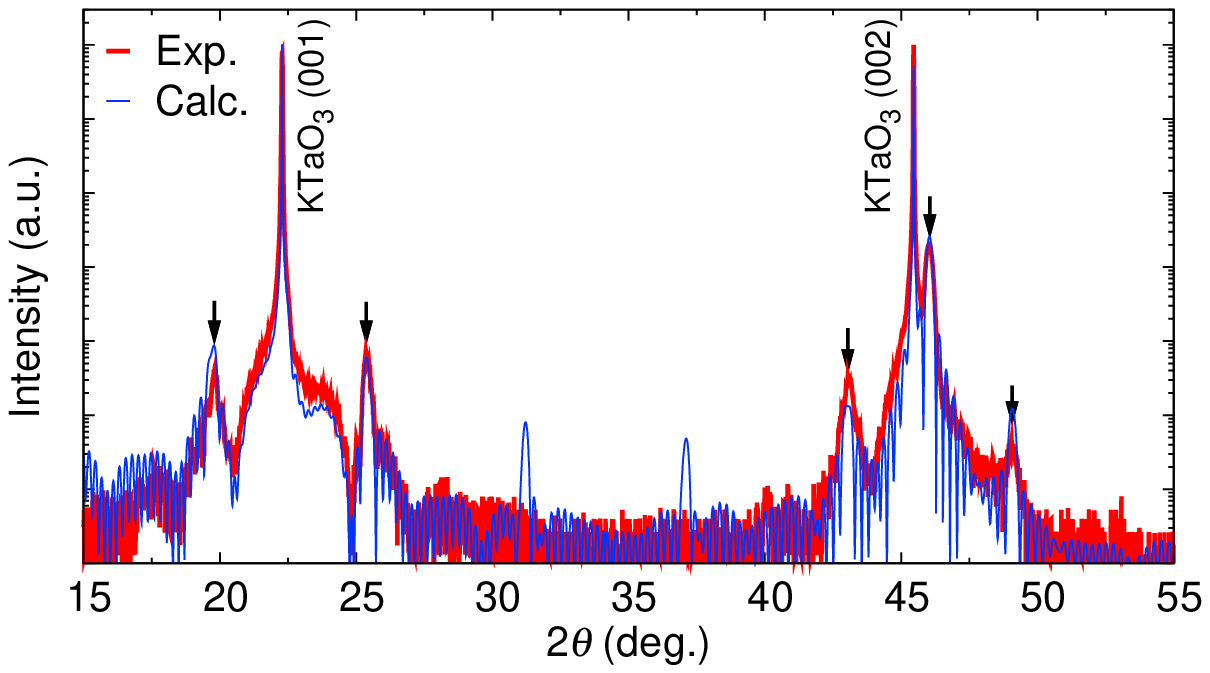}
\caption{
{\color[rgb]{0,0,0}Out-of-plane $\theta$--$2\theta$ XRD patterns of 
a film of 10[(SrMoO$_3$)$_4$/(SrTiO$_3$)$_4$] on 
the (001)-oriented KTaO$_3$ substrate. 
The blue line represents a dynamical theory diffraction 
calculation~\cite{TakahashiSS1995,YashiroSS2001} 
for the ideal structure of 10[(SrMoO$_3$)$_4$/(SrTiO$_3$)$_4$].
The fundamental and satellite peaks observed are indicated by arrows.}
}
\label{fig.s2}
\end{center}
\end{figure}

{\color[rgb]{0,0,0}
Figure~\ref{fig.s2} shows the $\theta$--$2\theta$ XRD scan of 
a (SrMoO$_3$)$_m$/(SrTiO$_3$)$_t$ film on the (001)-oriented KTaO$_3$ substrate. 
The film has ten repetitions of the $(m,t) = (4,4)$ structure, 10[(SrMoO$_3$)$_4$/(SrTiO$_3$)$_4$], 
as similar to the film on the SrTiO$_3$ substrate, 
shown in Fig.~4 in the main text. 
Fundamental and superlattice satellite peaks are clearly seen. 
The simulated pattern for the $(4,4)$ structure agrees with the experimental
results, suggesting that the quality of the film is good within 
the confirmation of the XRD level. 
The electrical resistivity $\rho$ of this film is shown in Fig.~5 of the main text. 
}
%

\section*{Magnified view of the STEM image of (SrMoO$_3$)$_4$/(SrTiO$_3$)$_4$}
\begin{figure}[h]
\begin{center}
 \includegraphics[width=0.55\textwidth]{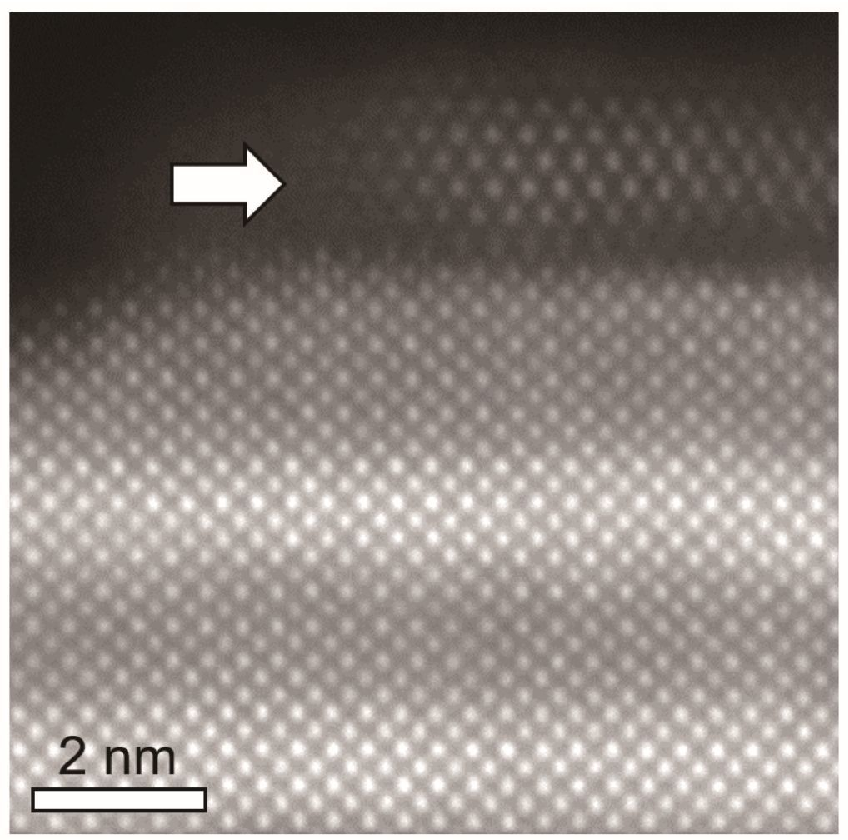}
\caption{
Magnified view of the high resolution HAADF-STEM image of 
the (SrMoO$_3$)$_4$/(SrTiO$_3$)$_4$ superlattice,
fabricated on the (001)-oriented SrTiO$_3$ substrate.
}
\label{fig.s3}
\end{center}
\end{figure}

In Fig.~\ref{fig.s3}, we present the magnified view of 
the high-angle annular dark-field (HAADF) STEM image 
{\color[rgb]{0,0,0}of
a superlattice film of 10[(SrMoO$_3$)$_4$/(SrTiO$_3$)$_4$] on 
the SrTiO$_3$ substrate. 
The sample is the same as that shown in Fig.~4 of the main text.}
We found another crystalline phase, as indicated by the arrow in Fig.~\ref{fig.s3}, 
in the surface of the SrMoO$_3$ film or interface between the SrMoO$_3$ and SrTiO$_3$ films.
The structure of this additional phase is different from the perovskite structure,
and the amount is minute,
which was clarified by increasing contrast of the image.
The presence of such a impurity may affect the crystallinity of SrMoO$_3$ 
around the heterointerface with SrTiO$_3$, 
resulting in interface roughing or intermixing disorder.
It is known that an impurity phase with the Mo$^{6+}$ 
valence state is often observed in the growth of 
SrMoO$_3$ films~\cite{WadatiPRB2014, Radetinac2016, Salg2019}. 
%


\end{document}